\documentclass[reprint,superscriptaddress,amsmath,amssymb,aps,prl]{revtex4-2}
\usepackage{amsmath, amsfonts}
\usepackage{float}
\usepackage{graphicx} 
\usepackage{dcolumn} 
\usepackage{bm} 
\usepackage{xr-hyper}
\begin{document}

\title{Microscopic Theory of Light-Induced Coherent Phonons Mediated by \\ Quantum Geometry}

\author{Jiaming Hu}
\affiliation{Center for Quantum Matter, School of Physics, Zhejiang University, Hangzhou 310058, China.}
\affiliation{Department of Materials Science and Engineering and Zhejiang Key Laboratory of 3D Micro/Nano Fabrication and Characterization, Westlake University, Hangzhou 310030, China.}

\author{Zhichao Guo}
\affiliation{Center for Quantum Matter, School of Physics, Zhejiang University, Hangzhou 310058, China.}

\author{Wenbin Li}
\email{liwenbin@westlake.edu.cn}
\affiliation{Department of Materials Science and Engineering and Zhejiang Key Laboratory of 3D Micro/Nano Fabrication and Characterization, Westlake University, Hangzhou 310030, China.}

\author{Hua Wang}
\email{daodaohw@zju.edu.cn}
\affiliation{Center for Quantum Matter, School of Physics, Zhejiang University, Hangzhou 310058, China.}

\author{Kai Chang}
\affiliation{Center for Quantum Matter, School of Physics, Zhejiang University, Hangzhou 310058, China.}

\date{\today}

\begin{abstract}
    Light-induced coherent phonon generation provides a powerful platform for ultrafast control of material properties. However, the microscopic theory and quantum geometric nature of this phenomenon remain underexplored. Here, we develop a fully quantum-mechanical framework based on Feynman diagrams to systematically describe the generation of coherent phonons by light. We identify a dominant second-order, double-resonant process in noncentrosymmetric semiconductors that efficiently couples light to both electronic and phononic excitations. Crucially, we uncover the quantum geometric origin, encoded in the electron-phonon coupling (EPC) shift vector and the EPC quantum geometric tensor. Applying our theory to ferroelectric BaTiO$_3$ and SnSe, we demonstrate the potential for light-induced modulation of ferroelectric polarization driven by coherent phonons. This work provides fundamental insights  for designing efficient optical control strategies for both coherent phonons and ferroelectric polarization.
\end{abstract}

\maketitle

\textit{Introduction}---Light-based manipulation has emerged as a powerful and versatile strategy for controlling lattice dynamics over the past few decades~\cite{guo2021recent,huang2010ferroelectric,wei2017photostriction,choi2009switchable}. Extensive research has particularly highlighted the feasibility of light-induced coherent phonons, revealing a wide range of applications, including Raman spectroscopy~\cite{tolles1977review,PhysRevResearch.5.L032016,nonadiabatic}, electronic structure renormalization~\cite{PhysRevX.15.021039}, solid-state ionic transport~\cite{pham2025dynamical}, electron-phonon coupling (EPC) detection~\cite{gerber2017femtosecond}, ultrafast ferroelectric polarization switching~\cite{subedi2015proposal,mankowsky2017ultrafast,istomin2007dynamics,cavalleri2006tracking,rubio2015ferroelectric,yang2018light,wang2017photo}, and multiferroic phase transition~\cite{zhou2021terahertz}. To unravel the underlying mechanisms, numerical simulations have been employed to provide microscopic insights into the coupled light–lattice dynamics~\cite{yang2024light,hu2024phonon}. Several theoretical studies, primarily based on phenomenological or simplified models, have also been conducted~\cite{chen2024ferroelectric,subedi2014theory,PhysRevB.45.768,merlin1997generating,10.1063/5.0188537,PhysRev.166.757}. However, direct coherent phonon excitation by light is challenging because it requires terahertz technology, which is generally less common and more difficult to implement than conventional mid-infrared or visible lasers. This highlights the importance of indirect and nonlinear excitation pathways. Furthermore, the microscopic theory of the complex dynamical effects arising from light-driven electronic excitations and the perturbation of EPC by optical fields remains underexplored.

To address this challenge, in this Letter, we adopt a rigorous Feynman diagrammatic approach to systematically describe light-induced coherent phonon generation within a fully quantum-mechanical framework. We identify the double-resonant process—resonant with both phonon and electron excitations—as a second-order nonlinear optical response and the dominant effect in noncentrosymmetric semiconductors. We further uncover its profound quantum geometric origin, governed by the EPC quantum geometric tensor and EPC shift vector. Finally, we apply our theory to first-principles calculations of ferroelectric bulk BaTiO$_3$ and two-dimensional SnSe, providing valuable insights for designing efficient coherent phonon generation schemes.


\textit{Coherent phonon generation.}---We begin by considering a set of harmonic $\Gamma$-point phonons described by the Hamiltonian $\hat{H}_{\rm ph} = {\sum\limits}_s\hbar\omega_s(\hat{b}^{\dagger}_s\hat{b}_s + \frac{1}{2})$, where $s$ denotes the phonon mode index, and $\hat{b}^{\dagger}_s$, $\hat{b}_s$ are the corresponding creation and annihilation operators. Next, we introduce a time-dependent phonon-generation kernel $\Gamma_s(t)$, which leads to an interaction operator initiated at $t=0$, $\hat{V}_s(t) = \theta(t)\Gamma_s(t)[\hat{b}^{\dagger}_s(t) + \hat{b}_s(t)]$, with $\hat{b}_s(t) = \hat{b}_s(0)e^{-i\omega_s t}$ and $\theta(t)$ being the step function. In general, $\Gamma_s(t)$ can represent any potential/interaction that couples to the phonon displacement $\hat{Q}_s = \bar{Q}_s[\hat{b}_s^\dagger(t) + \hat{b}_s(t)]$, where $\bar{Q}_s$ is the phonon displacement quantum. A typical example is the electronic component of EPC~\cite{giustino2017electron}.
. As detailed in Sec.~I~\cite{SI_info}, the interaction induces time evolution of the phonon ground state $|0_s\rangle$ via the operator $e^{i\hat{\alpha}_s(t)}|0_s\rangle$, defined as:
\begin{equation}\label{eq:alpha_s}
    \begin{aligned}
        \hat{\alpha}_s(t) &= \frac{1}{\hbar} \int_0^t \hat{b}_s^{\dagger}(t') \Gamma_s(t') \, dt' \\
        &= \frac{1}{\hbar} \int_0^t \hat{b}_s^{\dagger}(0) e^{i\omega_s t'} \int d\omega'\, \Gamma_s(\omega') e^{-i\omega' t'} \, dt'.
    \end{aligned}
\end{equation}
In the resonant limit $\omega' \to \omega_s$, this expression simplifies to $\hat{\alpha}_s(t) \to \hat{b}_s^{\dagger}(0)\frac{\Gamma_s(\omega_s) t}{\hbar}$, where the kernel $\Gamma_s(\omega_s)$ performs continuous positive work, coherently generating phonons with maximum efficiency. 

\begin{figure}[htp]
    \includegraphics[width=0.48 \textwidth]{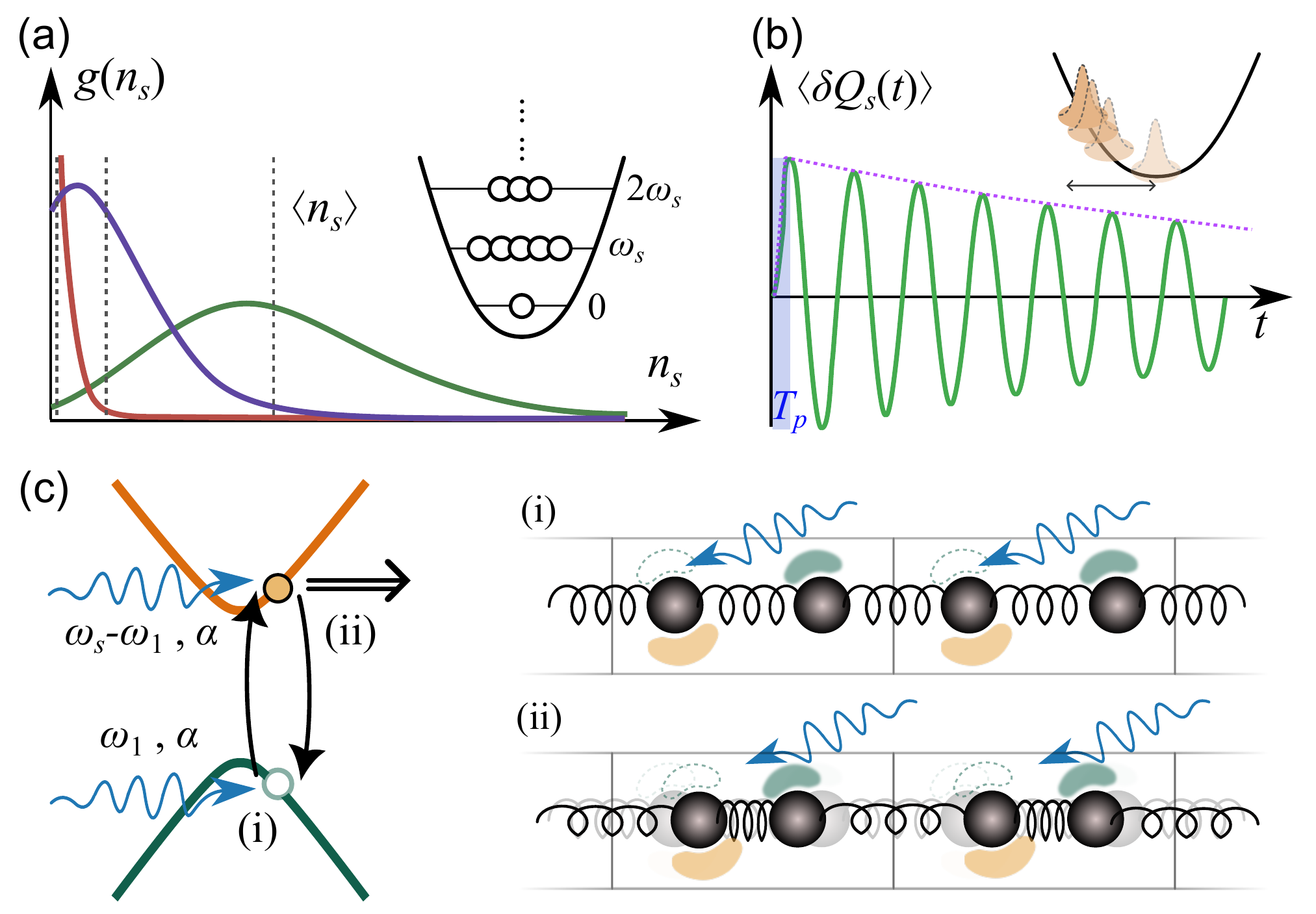}
    \caption{\label{fig:schematic_generation} \textbf{Schematic of light-induced coherent phonon generation.} (a) Phonon number distribution $g(n_s)$ of the coherent phonon state as discussed in Eq.\ref{eq:coherent_state}, with $|\Gamma_s(\omega_s) T_p / \hbar|=0.1,1,2$ drawn in red, purple and green, corresponding to different excitation strength. (b) Time evolution of the coherent phonon displacement $\langle \delta \hat{Q}_s(t) \rangle$, described by Eq.\ref{eq:delta_Q_s} with the decay due to decoherence. The blue shaded area marks the excitation of a femtosecond laser pulse, with pulse duration $T_p$ typically much shorter than the phonon period. (c) Schematic of phonon generation via EPC: (i) an electron is resonantly excited by light in frequency $\omega_1$, then (ii) relaxes by another light $\omega_s-\omega_1$ by transferring energy to a phonon through EPC. Yellow and green shaded regions illustrate electron clouds; the blue wavy line represents the incident light. }
\end{figure}

After a driving duration $T_p$, the ground state evolves into a coherent phonon state $|s,T_p\rangle$, given by:
\begin{equation}\label{eq:coherent_state}
    \begin{aligned}
        |s,T_p\rangle &= e^{i\hat{\alpha}_s(T_p)} |0_s\rangle \\
        &\propto \sum_{n_s=0}^{\infty} \frac{(i\Gamma_s(\omega_s)T_p/\hbar)^{n_s}}{\sqrt{n_s!}} |n_s\rangle,
    \end{aligned}
\end{equation}
which is a coherent superposition of $n_s$-phonon excitation states $|n_s\rangle$ ($n_s = 0, 1, 2, \ldots$). The probability distribution is given by $g(n_s) = e^{-|\Gamma_s(\omega_s) T_p / \hbar|^2}\frac{|\Gamma_s(\omega_s) T_p / \hbar|^{2n_s}}{n_s!}$, as schematized in Fig.~\ref{fig:schematic_generation}(a). The average phonon number is:
\begin{equation}
    \langle n_s \rangle = \frac{\langle s,T_p | \hat{b}^{\dagger}(t) \hat{b}(t) | s,T_p \rangle}{\langle s,T_p | s,T_p \rangle} = \left|\frac{\Gamma_s(\omega_s) T_p}{\hbar}\right|^2,
\end{equation}
which corresponds to the dashed lines in Fig.~\ref{fig:schematic_generation}(a). This coherent excitation gives rise to a nonzero phonon displacement:
\begin{equation}\label{eq:delta_Q_s}
    \begin{aligned}
        \langle \delta \hat{Q}_s(t) \rangle &= \frac{\langle s,T_p | \bar{Q}_s \left( \hat{b}_s^{\dagger}(t) + \hat{b}_s(t) \right) | s,T_p \rangle}{\langle s,T_p | s,T_p \rangle} \\
        &= 2\bar{Q}_s \frac{|\Gamma_s(\omega_s)| T_p}{\hbar} \cos(\omega_s t - \phi_s),
    \end{aligned}
\end{equation}
where $\phi_s$ is the phase of $i\Gamma_s(\omega_s)$. As a result, ions vibrate collectively along mode $s$ with amplitude $2\bar{Q}_s |\Gamma_s(\omega_s)| T_p / \hbar$, as illustrated in Fig.~\ref{fig:schematic_generation}(b). The key quantity determining the coherent phonon generation is the resonant component of the phonon-driving kernel $\Gamma_s(\omega_s)$, which is driven by optical electric fields $\bm{E}(\omega_{1,2})$ and can be perturbatively expanded as:
\begin{equation}
    \begin{aligned}
        \Gamma_s(\omega_s) &= \sum_{\alpha} \Gamma_s^{\alpha}(\omega_s; \omega_1) E^{\alpha}(\omega_1) \\
        &\quad + \sum_{\alpha\beta} \Gamma_s^{\alpha\beta}(\omega_s; \omega_1, \omega_2) E^{\alpha}(\omega_1) E^{\beta}(\omega_2) + o(|\bm{E}|^3),
    \end{aligned}
\end{equation}
where $\alpha, \beta$ label spatial directions and $\omega_{1,2}$ denote the frequencies of the applied optical fields.


\begin{figure}[htp]
    \includegraphics[width=0.48 \textwidth]{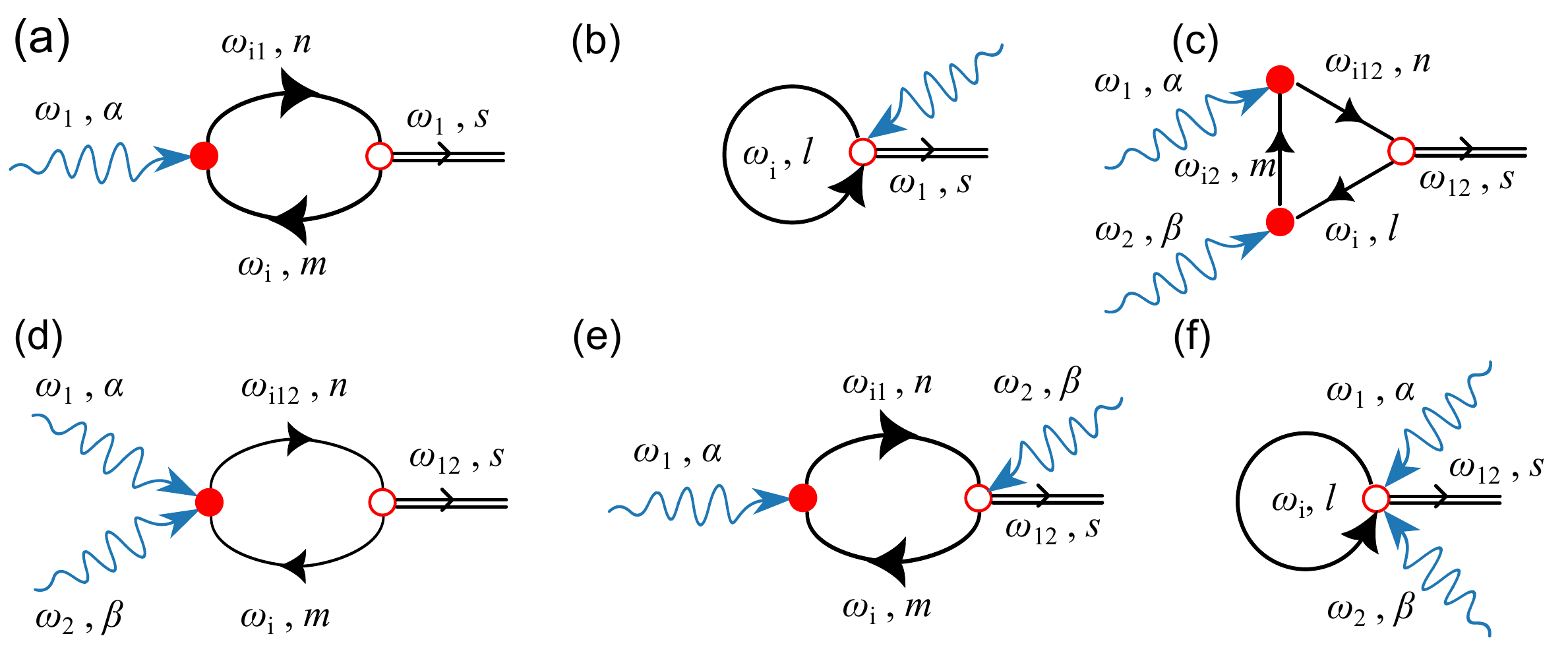}
    \caption{\label{fig:phonon_generation} \textbf{Feynman diagrams of light-induced phonon generation kernel.} (a,b) First- and (c-f) Second-order EPC-mediated kernel. The blue wavy arrow, black single arrow and black double arrow represent optical electric field, electron propagator and phonon propagator, respectively. The filled and hollow circle represents the electron-light and electron-phonon vertex, respectively. }
\end{figure}

\textit{Feynman diagram analysis.}---We calculate the phonon-driving kernel using the Feynman diagram approach developed in our previous work~\cite{PhNLO}. The first-order kernel $\Gamma^{\alpha}_s(\omega;\omega_1)$ originates from both direct and EPC-mediated processes. The direct contribution arises simply from the resonance between the photon and phonon frequencies.  
In the EPC-mediated diagrams, light perturbs electrons via inter-band excitation (Fig.~\ref{fig:phonon_generation}(a)) or intra-band drift (Fig.~\ref{fig:phonon_generation}(b)), which subsequently transfer energy to phonons via EPC. As derived in Sec.~I~\cite{SI_info}, the resulting contribution is:
\begin{equation}
    \Gamma^{\alpha}_s(\omega_s;\omega_1) = \delta(\omega_s - \omega_1) \times i e \int [d\bm{k}] \sum_{mn} \frac{f_{mn} g^s_{mn} h^{\alpha}_{nm}}{(\hbar\omega_1 + \epsilon_{mn}) \epsilon_{nm}}, 
\end{equation}
where $\delta(\omega_s - \omega_1)$ represents the conservation between $\omega_1$ and $\omega_s$. $g^s_{mn}$, $h^{\alpha}_{nm}$ are the $s$-mode EPC and $\alpha$-direction velocity (times $\hbar$) matrix elements between electronic bands $m,n$ at electronic wavevector $\bm{k}$ ($\bm{k}$ index is omitted for simplicity). The dipole matrix element $r^{\alpha}_{mn}$ is associated with inter-band velocity matrix element as $h^{\alpha}_{mn}=i\epsilon_{mn}r^{\alpha}_{mn}$~\cite{parker2019diagrammatic}. $\epsilon_{mn} = \epsilon_m - \epsilon_n$ is the electronic energy difference, $f_{mn} = f(\epsilon_m) - f(\epsilon_n)$ is the Fermi-Dirac distribution difference, and $\int[d\bm{k}]=\int d\bm{k}^d/(2\pi)^d$ denotes integration over the $d$-dimensional Brillouin zone. Under time-reversal symmetry (TRS), this kernel can be reformulated as:
\begin{equation}
    \begin{aligned}
        &\Gamma^{\alpha}_s(\omega_s;\omega_1) = \delta(\omega_s - \omega_1) \\
        &\times i \pi e \bar{Q}_s\hbar\omega_1 \int [d\bm{k}] \sum_{mn} f_{mn} \mathrm{Im}[\mathbb{G}^{s;\alpha}_{mn}] \delta(\hbar\omega_1 + \epsilon_{mn}),
    \end{aligned}
\end{equation}
where $\mathrm{Im}[\mathbb{G}^{s;\alpha}_{mn}]$, the imaginary part of the EPC quantum geometric tensor $\mathbb{G}^{s;\alpha}_{mn}$~\cite{PhNLO}, represents the covalent contribution to the Born effective charge~\cite{citro2023thouless,onoda2004topological}. It is important to note that $\Gamma^{\alpha}_s(\omega_s;\omega_1)$ is not included in the classical charge response of direct light-phonon coupling, where the covalent effective charge is calculated as $\int [d\bm{k}] \sum_{mn} f_{mn} \mathrm{Im}[\mathbb{G}^{s;\alpha}_{mn}]$. Instead, it represents a coherent superposition of all accessible quantum excitations selected by light frequency $\delta(\hbar\omega_1 + \epsilon_{mn})$. 
This term becomes resonant only when $\hbar\omega_s$ matches the electronic energy difference $\epsilon_{mn}$, contributing significantly in small-gap or metallic systems under terahertz light illumination, but less likely to resonate in wide-gap systems with mid-infrared or higher-frequency light. On the other hand


The second-order kernel $\Gamma^{\alpha\beta}_s(\omega_s;\omega_1,\omega_2)$ includes four distinct Feynman diagrams, as shown in Fig.~\ref{fig:phonon_generation}(c–f). However, 
only diagrams (c) and (e) are of particular interest, as they can simultaneously achieve electronic and phononic resonance ($\omega_1 + \omega_2 = \omega_s$). Specifically, diagram (c) captures two successive first-order optical transitions, characterized by light-perturbed electronic propagation. As derived in Sec.~II~\cite{SI_info}, it is calculated as: 
\begin{equation}
    \begin{aligned}
        &\Gamma^{\alpha\beta}_s(\omega;\omega_1,\omega_2)|_{c} 
        =
        \frac{-e^2}{\hbar\omega_1\hbar\omega_2}
        {\int}[d\bm{k}]{\sum\limits_{mnl}}
        \\ &{\times}
        g^s_{ln}h^{\alpha}_{nm}h^{\beta}_{ml}I_{lmn}(\omega_{2},\omega_{1}) + [\alpha,\omega_1{\leftrightarrow}\beta,\omega_2], 
    \end{aligned}
\end{equation}
where $I_{lmn}(\omega_{2},\omega_{1})$ captures electronic resonant structures as $I_{lmn}(\omega_{2},\omega_{1}) = \frac{(\hbar{\omega_{1}} + {\epsilon_{mn}})f_{lm} - (\hbar{\omega_{2}}-{\epsilon_{ml}})f_{mn}}{({\hbar}\omega_{2}-\epsilon_{ml})(\hbar\omega_{1}+\epsilon_{mn})(\hbar\omega_{12}-\epsilon_{nl})}$.
In diagram (e), as schematized in Fig.~\ref{fig:schematic_generation}(c), only one optical field invokes the electronic excitation while the other modifies the EPC vertex. This contribution is calculated as: 
\begin{equation}\label{eq:Gamma_2nd_origin}
    \begin{aligned}
        &\Gamma^{\alpha\beta}_s(\omega_s;\omega_1,\omega_2)\big|_{e} = \delta(\omega_s - \omega_1 - \omega_2) \\
        &\times \frac{-e^2}{\hbar\omega_1 \hbar\omega_2} \int [d\bm{k}] \sum_{mn} \left( \frac{f_{mn} g^{s;\beta}_{mn} h^{\alpha}_{nm}}{\hbar\omega_1 + \epsilon_{mn}} + \frac{f_{mn} g^{s;\alpha}_{mn} h^{\beta}_{nm}}{\hbar\omega_2 + \epsilon_{mn}} \right).
    \end{aligned}
\end{equation}
where $g^{s;\alpha}_{mn}$ refers to the matrix element of first-order covariant derivative of EPC operator. For simplicity, we further adopt the TRS and the wide-gap approximation $\hbar\omega_s \ll \epsilon_g$, where $\epsilon_g$ is the optical energy gap of the electron. We then keep only the zeroth order of $\hbar\omega_s/\epsilon_g$ as $\Gamma^{\alpha\beta}_s(\omega_s;\omega_1,\omega_s - \omega_1)|_{c,e} \approx \Gamma^{\alpha\beta}_s(\omega_s;\omega_1,-\omega_1)|_{c,e} + o(\hbar\omega_s/\epsilon_g)$, which leads to the total kernel of diagrams (c) and (e) as: 
\begin{equation}\label{eq:Gamma_2nd_final}
    \begin{aligned}
        \Gamma^{\alpha\beta}_s(\omega_s;\omega_1,&\omega_s-\omega_1)
        \approx 
        \pi e^2
        {\int}[d\bm{k}]{\sum\limits_{mnl}}\frac{f_{mn}}{\epsilon_{nm}}\delta(\hbar\omega_1-\epsilon_{nm})
        \\
        &{\times}
        [g^{s;\beta}_{mn} r^{\alpha}_{nm} + g^{s;\alpha}_{mn} r^{\beta}_{nm}
        \\
        &-\frac{h^{\beta}_{ml}g^s_{ln}r^{\alpha}_{nm}}{\hbar\omega_s+\epsilon_{ln}+i\eta}
        +
        \frac{r^{\beta}_{mn}g^s_{nl}h^{\alpha}_{lm}}{\hbar\omega_s+\epsilon_{nl}+i\eta}
        \\
        &-
        \frac{r^{\beta}_{mn}h^{\alpha}_{nl}g^s_{lm}}{\hbar\omega_s+\epsilon_{lm}+i\eta}
        +
        \frac{g^s_{ml}h^{\beta}_{ln}r^{\alpha}_{nm}}{\hbar\omega_s+\epsilon_{ml}+i\eta}], 
    \end{aligned}
\end{equation}
which confirms that even when light frequency $\omega_1 \gg \omega_s$, second-order processes can still provide a low-frequency driving force in the phonon frequency range, thus offering a key mechanism for coherent phonon generation induced by mid-infrared or higher-frequency light sources. Note that the approximation in Eq.~\ref{eq:Gamma_2nd_final} is with realistic consideration: in experiments, laser pulse typically has a Gaussian profile with a broadened intensity distribution around the central frequency $\omega_1$. As a result, one single laser source can effectively provide the necessary frequency component at $\omega_s - \omega_1\approx-\omega_1$ as long as $\omega_s \ll \omega_1$, making the coherent phonon excitation through Eq.~\ref{eq:Gamma_2nd_final} accessible.



\textit{Discussions.}---Eq.~\ref{eq:Gamma_2nd_final} captures the quasi-adiabatic contribution to EPC-induced phonon generation, which typically dominates in gapped systems~\cite{nonadiabatic,PhysRevB.82.165111}, where the phonon frequency is much smaller than the electronic energy gap. In this regime, the dynamical aspects of phonon vibrations are negligible compared to the timescale of electronic excitations. Therefore we only keep the influence of phonon frequency at the output side of the kernel as $\Gamma^{\alpha\beta}_s(\omega_s; \omega_1, - \omega_1)$, which is distinct from the fully adiabatic component $\Gamma^{\alpha\beta}_s(0; \omega_1, - \omega_1)$—the former is generally expected with higher efficiency and selectivity. Notably, the fully adiabatic effect is closely related to the ponderomotive force~\cite{PhysRevB.110.104301,huang2025universalphasetransitionsmatter} and optical forces~\cite{pimlott2025quantumgeometricinjectionshift}, acting as a phononic analogue of nonlinear photocurrent responses. In our framework, high-frequency light similarly generates a low-frequency effective “force” through nonlinear interactions, enabling resonant phonon generation. In both light-induced coherent phonon generation and BPVE, the breaking of inversion symmetry is essential for producing a nonzero spatially polarized response.



\textit{Quantum geometric origin.}---We further uncover the quantum geometric origin of the coherent phonon generation by applying a two-band approximation to Eq.~\ref{eq:Gamma_2nd_final} (Sec.~II~\cite{SI_info}) and focusing on the case of linearly polarized light. The second-order kernel then be expressed as:
\begin{equation}\label{eq:three_geo_terms}
    \begin{aligned}
        \Gamma^{\alpha\alpha}_s&(\omega_s;\omega_1,-\omega_1)|_{c+e}
        \approx - 2i\bar{Q}_s\pi e^2
        {\int}[d\bm{k}]{\sum\limits_{mn}}
        \\
        &{\times}f_{mn}\delta(\hbar\omega + \epsilon_{mn})\Big{(}{\rm Im}[\mathbb{G}^{s;\alpha}_{mn}]R^{s;\alpha}_{mn}
        \\
        &+ 
        {\rm Re}[\mathbb{G}^{s;\alpha}_{mn}]{\partial_{\alpha}}\ln(|g^s_{mn}|) 
        \\
        &+
        |r^{\alpha}_{nm}|^2(\frac{g^s_{mm}-g^s_{nn}}{\bar{Q}_s\hbar\omega_s})\Big{)}, 
    \end{aligned}
\end{equation}
where the three terms of Eq.~\ref{eq:three_geo_terms} correspond to distinct geometric effects of EPC: (i) EPC-induced charge shift: Governed by the EPC shift vector $R^{s;\beta}_{mn} = -\partial_{\beta}\arg(g^s_{mn}) + r^{\beta}_{mm} - r^{\beta}_{nn}$ with the shorthand $\partial_{\beta}\equiv\partial/\partial k^{\beta}$, which is a gauge-invariant quantity representing the EPC-induced shift of the electronic charge center~\cite{PhNLO, wang2024geodesicnaturequantizationshift}. This mechanism is analogous to the side-jump effect in the anomalous Hall effect~\cite{nagaosa2010anomalous}. (ii) Intrinsic EPC-induced deflection: Driven by the EPC quantum metric $\mathrm{Re}[\mathbb{G}^{s;\alpha}_{mn}]$, which determines the EPC-induced anomalous velocity~\cite{PhNLO}. This velocity is the EPC counterpart of the anomalous velocity induced by the electronic Berry curvature~\cite{xiao2010berry}, capturing the EPC-induced dispersion variation of electronic wavefunctions. (iii) EPC-induced skew scattering: Arising from asymmetric electron-phonon scattering of different bands, characterized by $(g^s_{nn} - g^s_{mm})$. This contribution is unrelated to inter-band coherence effects of EPC but rather stems from its chiral nature~\cite{nagaosa2010anomalous}. 

These mechanisms generate coherent phonons through distinct physical processes. In mechanism~(i), electrons are excited by light via the dynamic charge $e\mathrm{Im}[\mathbb{G}^{s;\alpha}_{mn}]$, which induces a shift in the electronic charge center quantified by $R^{s;\beta}_{nm}$. In mechanism~(ii), the electronic system is perturbed by light through the EPC dipole $\partial_{\alpha} \ln |g^s_{mn}|$, leading to a redistribution of the charge density dispersion characterized by $\mathrm{Re}[\mathbb{G}^{s;\alpha}_{mn}]$. Both the charge center shift and the variation in dispersion reflect modifications in the internal structure of the electronic system, which drive coherent phonon generation by altering the covalent bonding environment. In contrast, mechanism~(iii) involves purely intra-band EPC scattering $(g^s_{mm} - g^s_{nn}) / \bar{Q}_s$, manifesting as an asymmetric Hellmann--Feynman force induced by the polarization of the overall electronic system relative to the ionic background, which directly induces phonon excitation.



\textit{Application to photoferroelectric control.}---We apply our general theory to study one intriguing photoferroelectric effect, light modulation of ferroelectric (FE) polarization, by considering the polarization change induced by coherent phonon generation. To clearly elucidate the geometric mechanism, we first apply our theory to a dynamic Rice-Mele model, which serves as an effective two-site, one-dimensional (with direction labeled as $\mu$) representation for a broad class of ferroelectric materials (Sec.III~\cite{SI_info}). The staggered potential $\delta_m$ accounts for the influence of substrate or ionic species, while the static distortion $\gamma$ arises from in-plane FE polarization and modifies the inter-site hopping from $u$ to $u \pm \gamma u$. EPC is introduced via one in-plane optical phonon mode, which is denoted $\parallel$ with phonon displacement $Q_{\parallel}$. It modifies the inter-site distance, generating a dynamic distortion of $\pm vQ_{\parallel}$. The total Hamiltonian reads:
\begin{equation}\label{eq:RM_Hamiltonian}
    \hat{H}_k(Q_{\parallel}) = 
    \begin{bmatrix}
        \delta_m & V_k \\
        V_k^* & -\delta_m
    \end{bmatrix},
\end{equation}
where the off-diagonal term is $V_k = u\cos{\frac{ka}{2}} - i(\gamma u + vQ_{\parallel})\sin{\frac{ka}{2}}$, with $k$ the $\mu$-direction wavevector and $a$ the lattice constant. The conduction and valence bands are labeled $+,-$ respectively.

For conventional FE materials, we consider the weak-dispersion limit $\delta_m \gg u$ and the small-polarization limit $\gamma \ll 1$. Under these conditions, the EPC shift vector for the $\parallel$-mode, as the backbone of mechanism (i) of Eq.~\ref{eq:three_geo_terms}, simplifies to (see Sec.~III~\cite{SI_info}): $R^{\parallel;x}_{+-} \approx \mathrm{sign}(\delta_m)\frac{a\gamma}{2} \frac{\sin^2{(ka/2)}}{\cos^2{(ka/2)} + \gamma^2}$, which clearly exhibits polarity dependence on the inversion-symmetry-breaking parameters $\gamma$ and $\delta_m$. In contrast, EPC Berry curvature relates to dynamic charge hence does not depend on $\gamma$: $\mathrm{Im}[\mathbb{G}^{\parallel;\mu}_{+-}] \approx \mathrm{sign}(\delta_m)\frac{vua}{4\delta_m^2}\sin^2{\frac{ka}{2}}$, while EPC quantum metric that relevant to mechanism (ii) does exhibit $\gamma$-dependent behavior as: $ \mathrm{Re}[\mathbb{G}^{\parallel;\mu}_{+-}] = \frac{a\gamma}{2}\frac{vu}{4\delta_m^2}\sin{ka}$, indicating that the EPC-induced anomalous velocity is polarized by FE order. For mechanism (iii), the polarity is governed by the intra-band EPC difference $g^{\parallel}_{++} - g^{\parallel}_{--} = 2\bar{Q}_{\parallel} \frac{\gamma u v}{|\delta_m|}\sin^2{\left(\frac{ka}{2}\right)}$, which represents a FE-polarized EPC deformation field. Consequently, the second-order phonon-generation kernel driven by geometric mechanisms (i)--(iii) in Eq.~\ref{eq:three_geo_terms} scales as $\Gamma_{\parallel} \propto \gamma$, and changes sign under FE reversal $\gamma \to -\gamma$, indicating its potential ability to drive FE switching. 



\begin{figure}[htp]
    \includegraphics[width=0.48 \textwidth]{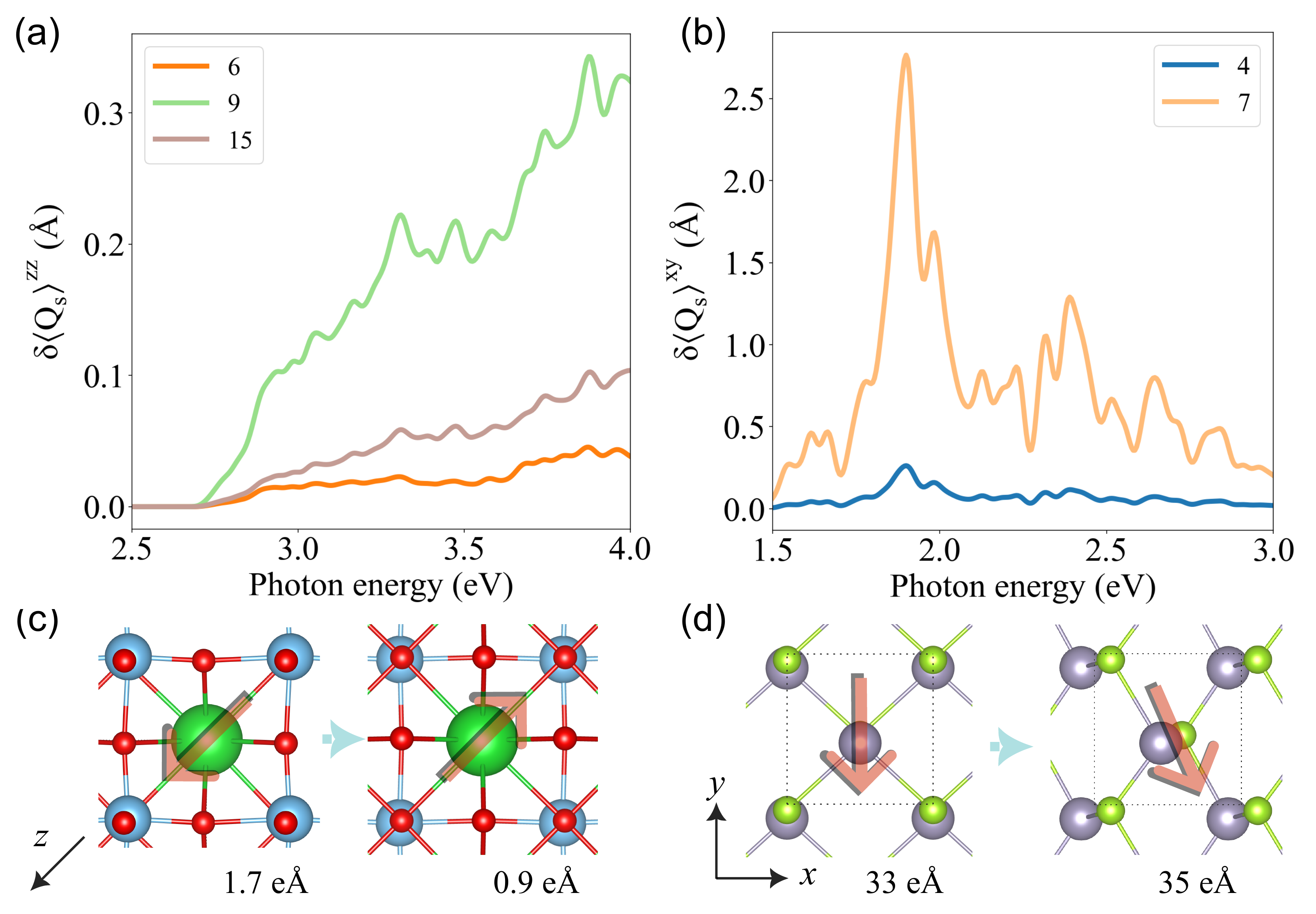}
    \caption{\label{fig:abinito_BTO} \textbf{First-principles calculation of light-induced coherent phonon generation.} Coherent vibration amplitudes of different phonon modes induced by (a) $z$-polarized light in rhombohedral BaTiO$_3$, and (b) $x$/$y$-polarized light in monolayer SnSe. The primary FE modes  correspond to the 9th mode in BaTiO$_3$ and the 7th mode in SnSe. Panels (c,d) illustrate the corresponding structural distortions caused by coherent phonon excitation, from left to right, for (c) BaTiO$_3$ under excitation at 3.8~eV and (d) SnSe under excitation at 1.8~eV. The polarization direction is indicated by arrows. }
\end{figure}
\textit{First-principles calculations.}---We then apply our theory to investigate laser-induced FE switching in representative three- and two-dimensional FE materials, namely rhombohedral BaTiO$_3$ and monolayer SnSe. The second-order kernel 
$\Gamma^{\alpha\beta}_s(\omega_s;\omega_1,-\omega_1)$, given by Eq.~\ref{eq:Gamma_2nd_final}, 
is obtained from first-principles calculations, with details provided in Sec.~IV~\cite{SI_info}. According to Eq.~\ref{eq:delta_Q_s}, the phonon vibration amplitude is estimated as $\langle \delta\hat{Q}_s \rangle^{\alpha\beta}(\omega_1) = \frac{2\bar{Q}_s}{\hbar} 
|\Gamma^{\alpha\beta}_s(\omega_s;\omega_1,-\omega_1)|E_0^2T_p $. For a typical laser setup with average optical field intensity 
$E_0 = 0.13$~V/\AA~and pulse duration $T_p = 10$~fs, selected results for dominant phonon modes in BaTiO$_3$ and SnSe are shown in 
Fig.~\ref{fig:abinito_BTO}(a,b). For linearly polarized light along $z$--direction, the 6th, 9th, and 15th modes of BaTiO$_3$ are predominantly excited, while for circularly polarized light in the $x\mbox{-}y$ plane, only the 4th and 7th modes are selectively excited. This mode selectivity, which stems from the material's symmetry, is discussed in detail in Sec.~IV~\cite{SI_info}. 

Next, we estimate the modulation strength of FE polarization using the same optical setup. For ultraviolet light with photon energy of 3.8~eV, the polarization of BaTiO$_3$ is reversed by $156\%$, switching from $-1.7$~e\AA~to $+0.9$~e\AA~along the $z$ direction, as illustrated in Fig.~\ref{fig:abinito_BTO}(c). 
For red light with photon energy of 1.8~eV, the in-plane polarization of SnSe is rotated by nearly $20^\circ$, as shown in Fig.~\ref{fig:abinito_BTO}(d). 
These results demonstrate that light can effectively modulate FE polarization in distinct ways depending on material and excitation conditions, thereby providing valuable guidance for designing efficient, non-invasive, light-driven FE switching techniques.

Finally, we emphasize that phonon modes contributions to polarization are intrinsically nonlinear: the total polarization change cannot be described as a simple summation of  individual mode contributions. As discussed in Sec.~IV~\cite{SI_info}, this nonlinear effect is negligible in BaTiO$_3$, but becomes critical in SnSe, reflecting their distinct dimensionalities and symmetries. This highlights the strong interplay between different modes in light-induced polarization, consistent with the previous theoretical proposals ~\cite{subedi2015proposal} and experimental studies~\cite{mankowsky2017ultrafast}, which indicate that light modulation of FE polarization is not governed solely by the FE phonon mode, but rather by its interaction with other phonon modes.

\textit{Conclusion.}---In this Letter, we developed a fully quantum-mechanical framework based on Feynman diagrams to systematically describe light-induced coherent phonon generation. Our analysis identifies a dominant second-order, double-resonant mechanism in noncentrosymmetric systems, which is simultaneously resonant with both electronic and phononic excitations. We demonstrate that this process has a fundamental quantum-geometric origin, encoded in the electron-phonon coupling (EPC) shift vector and the EPC quantum geometric tensor. This insight reveals how light drives phonon excitation by redistributing covalent electrons both internally and externally relative to the ions, thus providing a comprehensive understanding of this nonlinear process. We further performed first-principles calculations to 
explore light-induced modulation of ferroelectric polarization in BaTiO$_3$ and SnSe, highlighting the crucial interplay between ferroelectric modes and high-energy phonon modes. Our results emphasize the significant potential of this mechanism for optical control of ferroelectric order, establishing a theoretical foundation for efficient, noninvasive strategies to manipulate coherent phonons and engineer functional material properties.

\textit{Acknowledgment.}---The work of J.H. and W.L. is supported by the National Natural Science Foundation of China (NSFC) under Project No. 62374136. H.W. acknowledges the support from the NSFC under Grants Nos. 12522411, 12474240, and 12304049. K. C. acknowledges the support from the Strategic Priority Research Program of the Chinese Academy of Sciences (Grants Nos. XDB28000000 and XDB0460000), the NSFC under Grants Nos. 92265203 and 12488101, and the Innovation Program for Quantum Science and Technology under Grant No. 2024ZD0300104.

\bibliography{references}
\end{document}